\documentstyle[12pt]{article}
\title{ Inferring  degree of nonextensivity for generalized entropies} 
\author{{Ramandeep S. Johal\thanks{e-mail: 
raman\%phys@puniv.chd.nic.in}}\\  
{\it Department of Physics, Panjab University,}\\ 
{\it Chandigarh -160 014, India. }} 
\date{\today}
\begin{document}
\baselineskip 24pt
\def\be{\begin{equation}}
\def\ee{\end{equation}}
\def\ba{\begin{eqnarray}}
\def\ea{\end{eqnarray}}
\maketitle
\begin{abstract}
The purpose of this note is to argue  that degree of nonextensivity
as given by the Tsallis distribution obtained from maximum entropy 
principle has a different origin than the nonextensivity  
inferred from pseudo-additive property of Tsallis entropy. 
\end{abstract}
\newpage
Tsallis' nonextensive formalism of statistical mechanics has been 
designed to treat those systems which cannot be treated within
Boltzmann-Gibbs formalism owing to the presence of long-range
interactions, spatio-temporal complexity, fractal dynamics and so on. 
However, compared to the successful  number of applications of this 
formalism, there are relatively fewer papers which clarify the
theoretical foundations of the formalism \cite{1}. In this regard, it is
important to seek clear guidelines to propose any valid 
forms for generalized statistical mechanics \cite{2}. One such basis
for generlizing Shannon entropy to $q$-entropies (including Tsallis
entropy) 
seems to be provided by the connection of Tsallis formalism with
$q$-calculus or quantum groups.
In this paper, we will utilize a general form of Tsallis entropy
motivated by this connection and argue that the degree of nonextensivity
as manifested by the maximum entropy principle has a different
origin than the nonextensivity that is apparent in the  
pseudo-additive property of Tsallis entropy. 

It was proposed in \cite{3}
that Tsallis entropy generally written as 
\be
S^{T}_{q} = \frac{1-\sum_{i=1}^{W} (p_i)^q}{q-1}, \label{en1}
\ee
may be given by
\be
S^{T}_{q} = -\sum_{i=1}^{W} [a_i] p_i,
\ee
where 
\be
[a_i] = \frac{q^{a_i}-1}{q-1},\label{qn}
\ee
and
\be
a_i = \frac{q-1}{{\rm ln}\;q} {\rm ln}\;p_i .\label{ea}
\ee
In other words, the generalized bit-number can be written 
as Jackson's $q$-number. It was also argued that Eq. (\ref{ea})
is a tranformation which connects non-commutative differential
calculus to $q$-calculus. Although this makes Tsallis entropy
related to $q$-calculus, yet there is no justification
from $q$-theoretic arguments as to why the variable $a_i$
should depend on same $q$ parameter
as the  $q$-number $[a_i]$. It may well be that 
Tsallis entropy is a special case of an entropy function
where the $q$ in Eq. (\ref{ea}) coincides with the parameter
$q$ of Eq. (\ref{qn}). Motivated by this, we propose to work with
the entropy $S_q = -\sum_{i=1}^{W} [a_i] p_i$, where from now $a_i$
is given by
 \be
a_i = \frac{r-1}{{\rm ln}\;r} {\rm ln}\;p_i \label{eai}
\ee
Note that $r>0$. To see the role of parameter $r$, we maximize
$S_q$ under the generalized energy constraint
\be
\sum_{i}\varepsilon_i (q^{a_i}p_i) = U_q, \label{en3}
\ee
where $a_i$ is given by Eq. (\ref{eai}). Note that for $r=q$, the 
above constraint is equivalnet to $\sum_{i}\varepsilon_i p_{i}^{q} = U_q$,
which has been used earlier for the case of standard Tsallis entropy \cite{4}.
Now we study the variation of the function 
\be
\Phi = S_q - \alpha \sum_i p_i - \beta \sum_{i}\varepsilon_i (q^{a_i}p_i).
\ee
The probability distribution obtained by Lagrange multiplier method 
is given by
\be
p_i = \frac{\{1-(1-q){\beta}\varepsilon_i\}^{\frac{ {\rm ln}\;r}
{(1-r){\rm ln}\;q} } }{Z_{q,r}},
\label{en4}
\ee
where $Z_{q,r}$ is the partition function obtained from normalization
condition for $p_i$. Note that $q>0$, which is also significant
for concavity of entropy.
From Eq. (\ref{en4}), we see that when $r=q$, we get the usual 
Tsallis distribution    
\be
p_i = \frac{\{1-(1-q){\beta}\varepsilon_i\}^{1/(1-q)}}{Z_q},
\ee
Thus the exponent $1/(1-q)$ above actually  arises from the $a_i$
part (Eq. (\ref{eai})) of the definition of Tsallis entropy and not 
from the parameter of $q$-number (Eq. (\ref{qn})). In fact, it is
equally legitimate to work with more general form  \cite{5}
for $a_i$ given by  
\be
a_i = \frac{r-1}{{\rm ln}\;q} {\rm ln}\;p_i. 
\ee
Then the exponent of generalized distribution (Eq. (\ref{en4}))
is $1/(1-r)$.

Secondly, the pseudo-additive property of entropy $S_q = -\sum_i
[a_i]p_i$, where $a_i$ may depend either on $r$ or $q$,
\be
{S_q}(I+II)  = {S_q}(I) + {S_q}(II) + (1-q){S_q}(I){S_q}(II),\ee 
 follows directly from the $q$-additivity of $q$-numbers \cite{3}. 
This leads us to  state that the degree of nonextensivity
follows from different premises in the case of maximum
entropy principle and pseudo-additive property of Tsallis entropy,
respectively. 

\end{document}